\documentclass[twocolumn,amsmath,floatfix,prb,aps]{revtex4}

\usepackage{color}
\usepackage{amsmath}
\usepackage{pifont}   
\usepackage{graphicx} 
\usepackage{dcolumn}  
\usepackage{bm}       
\usepackage{amsfonts} 
\usepackage{amssymb}  
\usepackage{multirow} 
\usepackage{romannum}
\usepackage{siunitx}

\usepackage[english]{babel}
\usepackage[autostyle, english = american]{csquotes}
\MakeOuterQuote{"}

    \usepackage{braket}
    \usepackage{nicefrac}
    \usepackage{bm}
    \usepackage{physics}

    \renewcommand{\v}[1]{\bm{\mathrm{#1}}}

\begin{document}

\newcommand{\ba}{{\bf a}}
\newcommand{\BB}{{\bf b}}
\newcommand{\bd}{{\bf d}}
\newcommand{\br}{{\bf r}}
\newcommand{\bp}{{\bf p}}
\newcommand{\bk}{{\bf k}}
\newcommand{\bg}{{\bf g}}
\newcommand{\bj}{{\bf j}}
\newcommand{\bt}{{\bf t}}
\newcommand{\bv}{{\bf v}}
\newcommand{\bu}{{\bf u}}
\newcommand{\bq}{{\bf q}}
\newcommand{\bG}{{\bf G}}
\newcommand{\bP}{{\bf P}}
\newcommand{\bJ}{{\bf J}}
\newcommand{\bK}{{\bf K}}
\newcommand{\bL}{{\bf L}}
\newcommand{\bR}{{\bf R}}
\newcommand{\bS}{{\bf S}}
\newcommand{\bT}{{\bf T}}
\newcommand{\bQ}{{\bf Q}}
\newcommand{\bA}{{\bf A}}
\newcommand{\bH}{{\bf H}}

\newcommand{\bdel}{\boldsymbol{\delta}}
\newcommand{\bsig}{\boldsymbol{\sigma}}
\newcommand{\beps}{\boldsymbol{\epsilon}}
\newcommand{\bnu}{\boldsymbol{\nu}}
\newcommand{\bnab}{\boldsymbol{\nabla}}
\newcommand{\bGam}{\boldsymbol{\Gamma}}

\newcommand{\bgt}{\tilde{\bf g}}

\newcommand{\brh}{\hat{\bf r}}
\newcommand{\bph}{\hat{\bf p}}

\title{A general relation between stacking order and Chern index: a topological map of minimally twisted bilayer graphene}

\author{S. Theil$^1$}
\author{M. Fleischmann$^1$}
\author{R. Gupta$^2$}
\author{F. Wullschl\"ager$^3$}
\author{S. Sharma$^4$}
\author{B. Meyer$^3$}
\author{S. Shallcross$^4$}
\affiliation{1 Lehrstuhl f\"ur Theoretische Festk\"orperphysik, Staudtstr. 7-B2, 91058 Erlangen, Germany}
\affiliation{2 H. H. Wills Physics Laboratory, University of Bristol,Tyndall Avenue, Bristol BS8 1TL, United Kingdom}
\affiliation{3 Interdisciplinary Center for Molecular Materials (ICMM) and Computer-Chemistry-Center (CCC), Friedrich-Alexander-Universit\"at Erlangen-N\"urnberg (FAU), N\"agelsbachstra{\ss}e~25, 91052 Erlangen, Germany}
\affiliation{4 Max-Born-Institute for Non-linear Optics and Short Pulse Spectroscopy, Max-Born Strasse 2A, 12489 Berlin, Germany}

\date{\today}

\begin{abstract}
We derive a general relation between the stacking vector $\bu$ describing the relative shift of two layers of bilayer graphene and the Chern index. We find $C = \nu - \text{sign}\left(|V_{AB}|-|V_{BA}|\right)$, where $\nu$ is a valley index and $|V_{\alpha\beta}|$ the absolute value of stacking potentials that depend on $\bu$ and that uniquely determine the interlayer interaction; AA stacking plays no role in the topological character. With this expression we show that while ideal and relaxed minimally twisted bilayer graphene appear so distinct as to be almost different materials, their Chern index maps are, remarkably, identical. The topological physics of this material is thus strongly robust to lattice relaxations.
\end{abstract}

\maketitle

\section{Introduction}

Ideal and atomically relaxed twist bilayer graphene are, in the small angle regime, essentially different materials\cite{dai_twisted_2016,jain_structure_2016,
yoo_atomic_2019,gargiulo_structural_2017,nam_lattice_2017}. While the ideal lattice geometry is that of a moir\'e, for $\theta < 1^\circ$ the material relaxes ("reconstructs"\cite{yoo_atomic_2019}) into domains of AB and BA stacking bounded by pure screw partial dislocations\cite{Alden2013,butz14}. Evidently, a moir\'e encompassing equally all stacking types and an ordered structure of AB and BA domains are very different material systems. The remarkable electronic properties of the graphene twist bilayer have, however, predominately been established for the ideal geometry\cite{Shallcross2010,shall13,tram10,Bistritzer2011a,Weck16,san-jose_helical_2013} and a natural question is therefore how the rich electronic physics of the graphene moir\'e is impacted by the profound lattice relaxation that occurs at small angles\cite{nam_lattice_2017,lucignano_crucial_2019,angeli_emergent_2018}.

AB and BA stacked bilayer graphene have different valley Chern numbers, generating a pair of topologically protected states with valley momentum locking at the domain walls of regions of AB and BA stacking. In the ordered network of AB and BA domains that constitute minimally twisted bilayer graphene these one dimensional states lead to a "helical network" of valley-momentum locked states\cite{Huang2018,rickhaus_transport_2018}, and a remarkable electrically controllable and complete nesting of the Fermi surface\cite{fleischmann_perfect_2020}, with a correspondingly rich magneto-transport that is only beginning to be explored\cite{rickhaus_transport_2018,beule2020aharonovbohm}. In this paper we will ask the inverse question to that posed above: can such a network of one dimensional states be found in the moir\'e as well as the dislocation network? 

{\it A priori}, this would appear unlikely. It implies that the topological character of a material consisting of an ordered mosaic of AB and BA domains be identical to the smooth stacking modulation of the moir\'e. Remarkably, as we show here, the moir\'e and the partial dislocation network have identical topological character, in the sense that the spatial dependence of the valley Chern number is indistinguishable between these two systems. This represents the first example of a property of the twist bilayer fully robust to lattice relaxation and suggests (i) that the helical network will survive at twist angles when the relaxation to a dislocation network is incomplete, and (ii) that in Dirac-Weyl materials for which the energetic balance of in-plane strain and interlayer stacking energy may not favor reconstruction to a dislocation network, the physics of the "helical network" may nevertheless be found.

Our approach will be to generalize the widely known fact that AB and BA stacked bilayer graphene have different valley Chern numbers to a statement concerning an arbitrary stacking vector and the corresponding Chern index. Employing the fact that, under quite general assumptions, the interlayer interaction in bilayer graphene can be represented by three unique "stacking potentials" (corresponding to the three high symmetry stacking types of AB, BA, and AA stacking), we demonstrate that the valley Chern index $C$ depends only on the sign of the difference of the AB and BA potentials as 

\begin{equation}
C = \nu - \text{sign}\left(|V_{AB}|-|V_{BA}|\right)
\label{XX}
\end{equation}
with $\nu=\pm1$ an index labeling the conjugate K valleys. An intervening metallic state is required at a topological phase transition, and we show that the stacking phase diagram of bilayer graphene contains "permanent metal lines" at which the system remains metallic for any interlayer bias, and that these lines exactly correspond to the stacking vectors at which the valley Chern index changes (Sec. IV). We then numerically investigate the veracity of Eq.~\eqref{XX} through a series of artificial domain walls for which $C$ is predicted to be different or identical, as well as considering the case of one dimensional smooth stacking orders looking for bound states associated with a sign change of $|V_{AB}|-|V_{BA}|$ (Sec. VI). Finally, we show (Sec. VII) that employing Eq.~\eqref{XX} the topological character of both ideal and relaxed twist bilayer graphene is identical.

\section{Effective Hamiltonian theory}
\label{effective}

In Ref.~\onlinecite{Rost2019} it was shown that the tight-binding Hamiltonian exactly maps onto the following continuum Hamiltonian

\begin{equation}
[H(\br,\v{p})]_{\alpha \beta} = \frac{1}{A_{\mathrm{UC}}} \sum_{\v{G}_j} [M_{j}]_{\alpha\beta} \:\eta_{\alpha\beta} (\br,\v{K}_j+ \v{p})
\label{simplified_hamiltonian}
\end{equation}
where $\bG_j$ are the reciprocal lattice vectors and the sum thus represents the translation group of the expansion point $\bK_j$; the function $\eta_{\alpha\beta}$ is the Fourier transform of an envelope function describing the tight-binding matrix elements between $\br$ and $\br+\bdel$, $\eta_{\alpha\beta}(\br,\bq) = \int dr\, e^{i\bq.\bdel} t_{\alpha\beta}(\br,\bdel)$. The "M matrices" are given by

\begin{equation}
[M_j]_{\alpha\beta} = e^{\mathrm{i}\v{G}_j\cdot(\bnu_\alpha^n-  \bnu_\beta^m)}
\end{equation}
and encode through a mixed space representation the lattice and basis of the high symmetry system. For further details we refer the reader to Ref.~\onlinecite{Rost2019} as well as several applications of the method: to minimally twisted bilayer graphene\cite{fleischmann_perfect_2020}, partial dislocation networks\cite{kiss15,shall17,Weckbecker2019}, and in-plane deformation fields\cite{gupta19,gupta19a}.

The layer diagonal blocks of Eq.~\eqref{simplified_hamiltonian} can be Taylor expanded for slow deformation to yield the exact single layer tight-binding Hamiltonian plus deformation corrections expressed through (at lowest order) the pseudo-gauge and spatial variation of the Fermi velocity tensor\cite{gupta19}. For systems with both intra- and interlayer (stacking) deformations the electronic structure is dominated by the latter\cite{fleischmann_perfect_2020}, and so we will not include in-plane effective fields here. A general form for the interlayer interaction\cite{Rost2019} is given by

\begin{equation}
    [S(\v r,\v p)]_{\alpha\beta} = \frac{1}{A_{\mathrm{UC}}} \sum_j       [M_j]_{\alpha\beta}\:e^{-\mathrm{i}\Delta \v{u}(\v{r})\cdot          \v{G}_j}\:t(\v{K}_j+\v{p})
    \label{interlayer_coupling}
\end{equation}
where $\Delta \bu(\br)$ is a deformation field describing a local shift of the two layers by $\Delta \bu$ at $\br$.
The $C_3$ symmetry of graphene demands that each star of the translation group of momentum boosts encoded in the above equation is described by the same 3 "M matrices":

\begin{equation}
M_0=\begin{pmatrix}
1&1\\
1&1
\end{pmatrix},\qquad
M_\pm=\begin{pmatrix}
1&e^{\pm 2\pi\mathrm{i}/3}\\
e^{\mp 2\pi \mathrm{i}/3} & 1
\end{pmatrix}
\end{equation}
and for this reason the interlayer interaction can therefore be expressed as a sum of 3 distinct parts. The most convenient way in which the interlayer interaction can be decomposed is then in terms of the three stacking potentials associated with AB, BA and AA stacking, which have off-diagonal matrix structure $\sigma_+$, $\sigma_-$, and $\sigma_0$ respectively.

We thus have a general form for the Hamiltonian of bilayer graphene with arbitrary interlayer stacking

\begin{equation}
H=\begin{pmatrix}
    \Delta & \nu p_x-\mathrm{i}p_y & V_\mathrm{AA} & V_\mathrm{AB}\\
    \nu p_x+\mathrm{i}p_y & \Delta & V_\mathrm{BA} & V_\mathrm{AA}\\
    V_\mathrm{AA}^* & V_\mathrm{BA}^* & -\Delta & \nu p_x-\mathrm{i}p_y\\
    V_\mathrm{AB}^* & V_\mathrm{AA}^* & \nu p_x+\mathrm{i}p_y & -\Delta
\end{pmatrix}
\label{full}
\end{equation}
where we have truncated the layer-diagonal blocks at linear order, which is convenient for the analytical work which follows. The diagonal blocks are thus Dirac-Weyl operators with valley index $\nu=\pm 1$, interlayer bias $\Delta$, and the Fermi velocity set to unity. The interlayer potentials $V_{AB}$, $V_{BA}$, and $V_{AA}$ can be obtained from Eq.~\eqref{interlayer_coupling}.

\section{From stacking order to Chern index}
\label{valleyChernSection}

\begin{figure}
\centering
\includegraphics[width=0.4\textwidth]{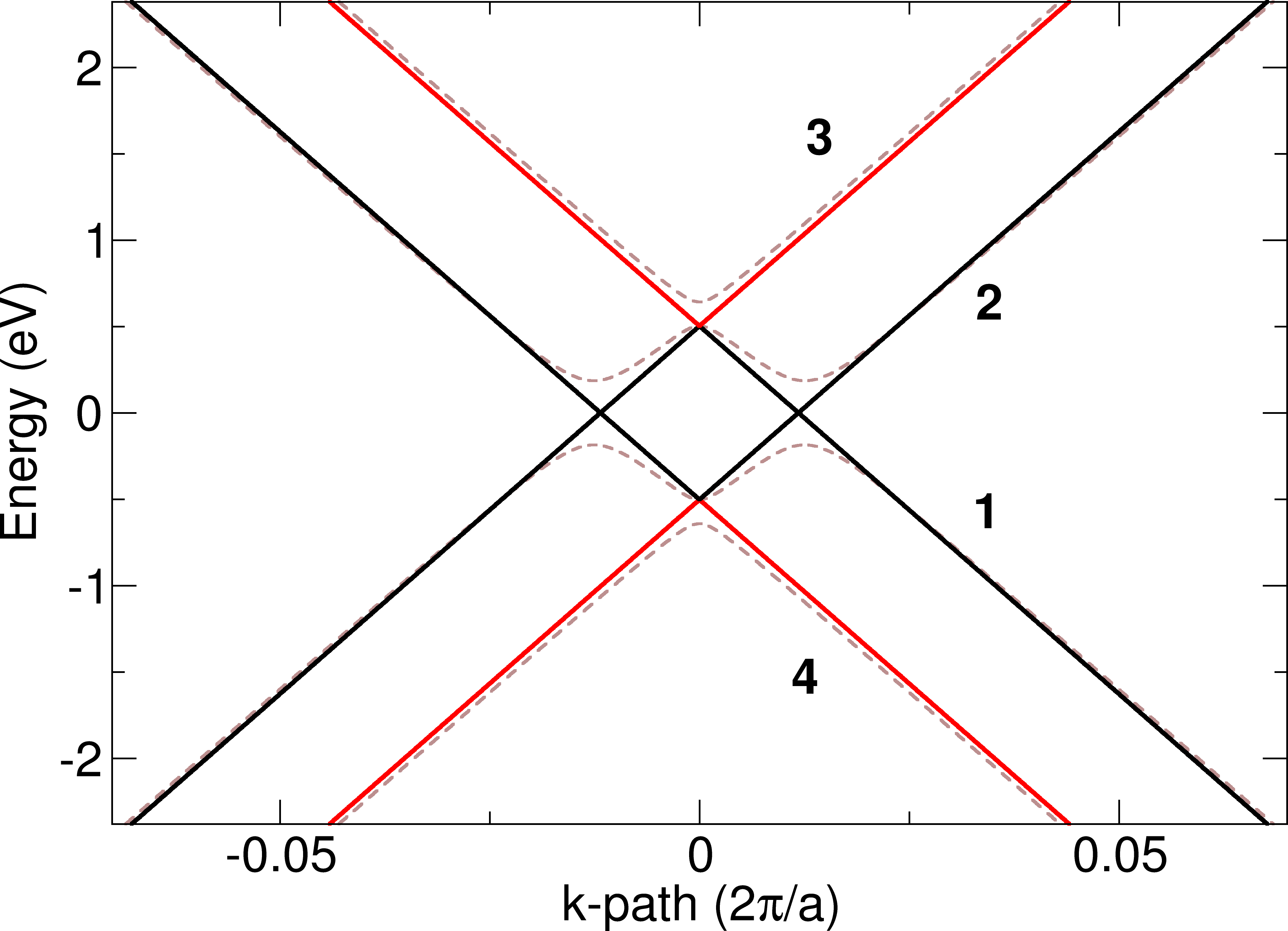}
\caption{Full lines: band structure of AB stacked bilayer graphene with interlayer bias but the interlayer interaction switched off. Broken lines: band structure of AB stacked bilayer graphene. The single layer bands indicated by the numbers 1-4 are employed as a basis in the calculations of Sec.~\ref{valleyChernSection} and Sec.~\ref{metal}.}
\label{fig0}
\end{figure}

Analytical calculation of the Berry curvature of Eq.~\eqref{full} would, for an arbitrary stacking, represent a formidable task. To simplify this we break the full Hamiltonian into two sub-systems: a low energy sector spanned by the single layer states labeled 1 and 2 in Fig.~\ref{fig0} and a high energy sector spanned by states 3 and 4. The two basis sets are therefore

\begin{equation}
\ket{\Psi_1}=\frac{1}{\sqrt{2}}\begin{pmatrix}
1\\
-\nu e^{\mathrm{i}\nu\phi}\\
0\\
0
\end{pmatrix}\qquad
\ket{\Psi_2}=\frac{1}{\sqrt{2}}\begin{pmatrix}
0\\
0\\
1\\
+\nu e^{\mathrm{i}\nu\phi}
\end{pmatrix}
\end{equation}
for the low energy sector, and 
\begin{equation}
\ket{\Psi_3}=\frac{1}{\sqrt{2}}\begin{pmatrix}
1\\
+\nu e^{\mathrm{i}\nu\phi}\\
0\\
0
\end{pmatrix}\qquad
\ket{\Psi_4}=\frac{1}{\sqrt{2}}\begin{pmatrix}
0\\
0\\
1\\
-\nu e^{\mathrm{i}\nu\phi}
\end{pmatrix}
\end{equation}
for the high energy sector. In these expressions \(\phi=\arctan(k_y/k_x)\) is the polar angle of the momentum. The justification for decomposing the full Hamiltonian in this way is that the Berry curvature will be associated with those parts of momentum space that, when the interlayer interaction is tuned to zero, have degenerate states. This is the physics captured by the low energy sector described by states 1 and 2. In calculating the Berry curvature for AB and BA stacking Zhang {\it et al}.\cite{zhang_valley_2013} employed an alternative basis of states 1 and 4, calculating the Berry curvature deep in the valence band. We find that for the case of a general stacking this leads to an erroneous AA contribution to the topological invariant; apparently for the more general case a careful treatment of the low energy bands becomes important. As we will show, our result reproduces as a limit those of Ref.~\onlinecite{zhang_valley_2013}.

The low energy Hamiltonian in the basis of states 1 and 2 is

\begin{equation}
H^{\mathrm{low}} = \begin{pmatrix}
\Delta-|k| & O\\
O^\ast &-\Delta+|k|
\end{pmatrix}
\end{equation}
while the high energy Hamiltonian in the basis of states 3 and 4 is

\begin{equation}
H^{\mathrm{high}}=
\begin{pmatrix}
\Delta+|k| & -O \\
-O^\ast &-\Delta-|k|
\end{pmatrix}
\end{equation}
where the off-diagonal elements are given by

\begin{equation}
O=\frac{\nu}{2}(V_\mathrm{AB}e^{\mathrm{i}\nu\phi} - V_\mathrm{BA}e^{-\mathrm{i}\nu\phi}):=\abs{a}e^{\mathrm{i}\theta}
\label{apot}
\end{equation}
The eigenvalues of these Hamiltonians are given by

\begin{equation}
E^{\mathrm{low}}= \pm\sqrt{(\Delta-\abs{k})^2+|O|^2}=:\pm  \xi,
\end{equation}
and
\begin{equation}
E^{\mathrm{high}}=\pm \sqrt{(\Delta+\abs{k})^2+|O|^2}=:\pm  \zeta
\end{equation}
with the eigenvectors given by

\begin{equation}
v^{\mathrm{low}}_\pm=\frac{1}{\sqrt{2}}\begin{pmatrix}
\sqrt{1\pm \frac{\Delta-\abs{k}}{\xi}}\\
\pm\sqrt{1\mp \frac{\Delta-\abs{k}}{\xi}}e^{-\mathrm{i}\theta}
\end{pmatrix}=:
\begin{pmatrix}
c_\pm\\
\pm c_\mp e^{-\mathrm{i}\theta}
\end{pmatrix}
\end{equation}

and

\begin{equation}
v^{\mathrm{high}}_\pm=\frac{1}{\sqrt{2}}\begin{pmatrix}
\sqrt{1\pm \frac{\Delta+\abs{k}}{\zeta}}\\
\pm\sqrt{1\mp \frac{\Delta+\abs{k}}{\zeta}}e^{-\mathrm{i}\theta}
\end{pmatrix}=:
\begin{pmatrix}
d_\pm\\
\pm d_\mp e^{-\mathrm{i}\theta}
\end{pmatrix}
\end{equation}

From these we can then reconstruct the wave functions in the original layer-sublattice space $\Phi^{\mathrm{low}}_\pm=v^{\mathrm{low}}_{\pm,1}\ket{\Psi_1} + v^{\mathrm{low}}_{\pm,2}\ket{\Psi_2}$ and $\Phi^{\mathrm{high}}_\pm=v^{\mathrm{high}}_{\pm,1}\ket{\Psi_3} + v^{\mathrm{high}}_{\pm,2}\ket{\Psi_4}$, and then determine the Berry connection $A^{\mathrm{low/high}}_\pm = -\mathrm{i}\braket{\Phi^{\mathrm{low/high}}_\pm}{\partial_\phi\Phi^{\mathrm{low/high}}_\pm}$, finding

\begin{equation}
A_\pm^{\mathrm{low}}=\frac{\nu}{2}(1-c_\mp^2\theta')
\end{equation}
for the low energy sector and

\begin{equation}
A_\pm^\mathrm{high}=\frac{\nu}{2}(1-d_\mp^2\theta')
\end{equation}
for the high energy sector, where $\theta'=\partial_\phi\theta$.
We must now sum over occupied states \(A^{\mathrm{low}}_-\) and \(A^{\mathrm{high}}_-\) to give

\begin{figure}
\centering
\includegraphics[width=0.49\textwidth]{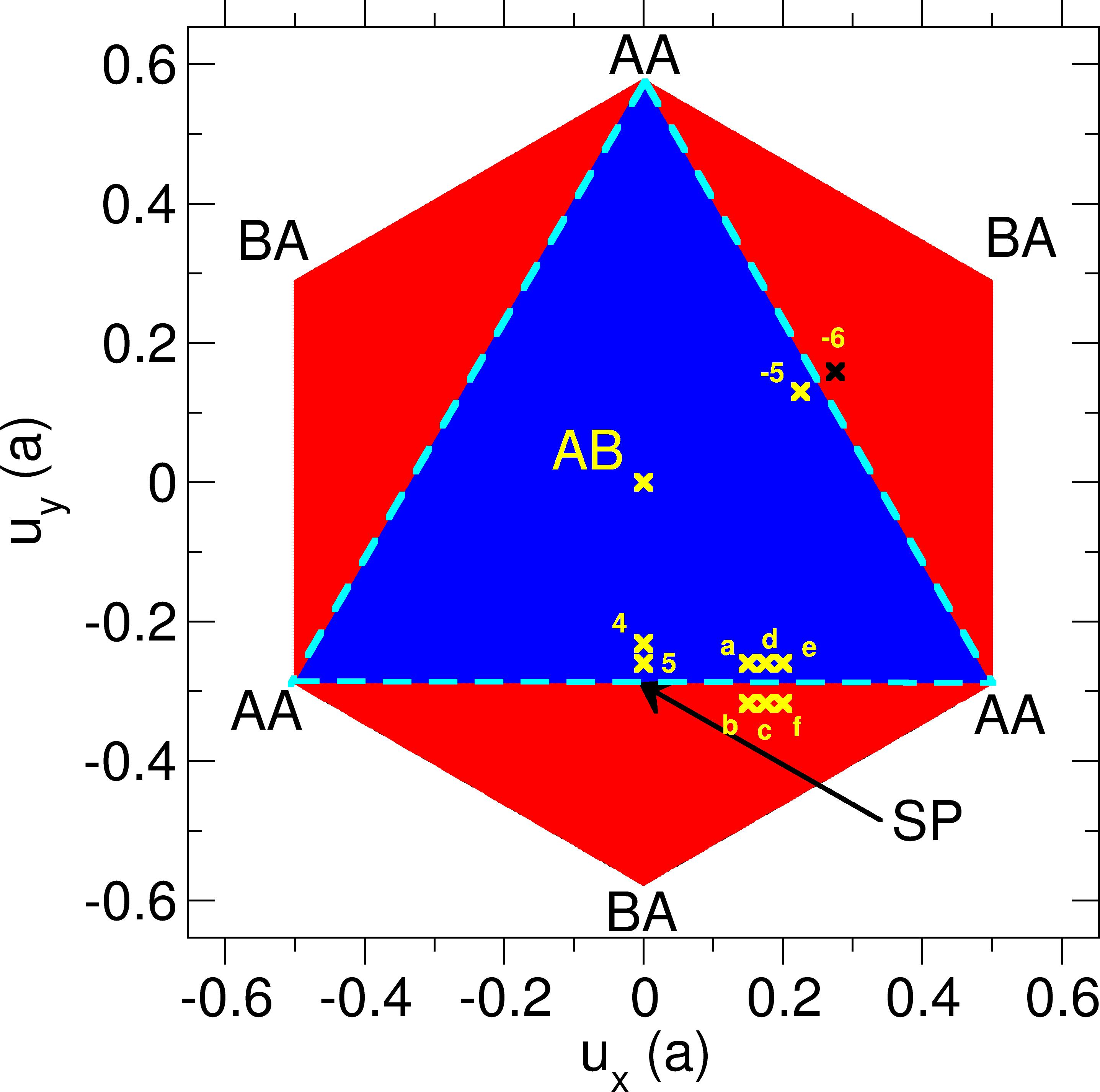}
\caption{Winding number phase diagram of bilayer graphene as a function of the relative shift of the two layers $\bu = (u_x,u_y)$ (measured in units of the lattice constant $a$); red denotes a winding number of -1 and blue a winding number of +1. AB stacking corresponds to $\bu ={\bf 0}$, the centre of the diagram, with BA and AA on the vertices. The stacking type SP is that found at a partial dislocation core and lies equidistant between AB and BA.
The dashed lines are the "metal lines" on which the system remains metallic irrespective of the magnitude of the interlayer potential. The labeled crosses are the stacking vectors used to construct artificial domain walls to probe bound states associated with changing winding number; each panel in Fig.~\ref{fig2} corresponds to a pair of stacking vectors.}
\label{fig1}
\end{figure}

\begin{equation}
A=A_-^\mathrm{low} + A_-^\mathrm{high} =  \nu -\nu\theta' -\frac{1}{2}\left(\frac{\Delta - \abs{k}}{\xi} + \frac{\Delta + \abs{k}}{\zeta}\right)
\end{equation}

In the limit of large momentum, we can neglect \(\Delta\) in the above expression and both \(\xi\) and \(\zeta\) reduce to \(\abs{k}\), so thus the bracketed term vanishes.
After integrating around a fixed \(\abs{k}\) path, we arrive at an expression for the Chern number given by

\begin{equation}
C=\nu-\frac{\theta(2\pi)-\theta(0)}{2\pi}
\label{trueChern}
\end{equation}
which depends only on the valley index $\nu=\pm 1$ and the winding number of \(O\), Eq.~\eqref{apot}. Expressing the generally complex potentials $V_{AB}$ and $V_{BA}$ as their absolute value and phase, the equation for $O$ can be written in polar coordinates in the complex plane as

\begin{eqnarray}
O&=&-\frac{\nu}{2}\left[\abs{V_\mathrm{AB}}e^{\mathrm{i}(\theta_\mathrm{AB}+\nu\phi)} - \abs{V_\mathrm{BA}}e^{\mathrm{i}(\theta_\mathrm{BA}- \nu \phi)}\right]\\
&=&-\frac{\nu}{2}e^{\mathrm{i}\frac{\theta_\mathrm{AB}-\theta_\mathrm{BA}}{2}}\Big[\left(\abs{V_\mathrm{AB}}-\abs{V_\mathrm{BA}}\right)\cos\phi' \nonumber \\
& &+                 \mathrm{i}\left(\abs{V_\mathrm{AB}}+\abs{V_\mathrm{BA}}\right)\sin\phi'\Big]
\end{eqnarray}
and we see that if \(|V_\mathrm{AB}|>|V_\mathrm{BA}|\), the ellipse turns counter clockwise and the winding number is plus one while if
 \(|V_\mathrm{AB}|<|V_\mathrm{BA}|\), the relative signs of the sine and cosine terms are different, and the winding number is minus one. The Chern number for arbitrary stacking is therefore

\begin{equation}
C = \nu - \text{sign}\left(|V_{AB}|-|V_{BA}|\right)
\label{C}
\end{equation}
where the potentials $V_{AB}$ and $V_{BA}$ are related to the stacking through Eq.~\eqref{interlayer_coupling}. Evidently, this result reduces to the correct form of the Chern index for AB (0 and 2 for the K and K' valleys respectivey) and BA (2 and 0 for the K and K' valleys respectively) derived in Ref.~\onlinecite{zhang_valley_2013}.

\section{Metallic lines in the stacking phase diagram}
\label{metal}

The phase diagram of winding number versus stacking vector is shown in Fig.~\ref{fig1}. In principle one can move through this phase diagram by sliding two layers of graphene and so cross a boundary separating distinct topological invariants. On such a boundary the gap must close irrespective of the magnitude of the interlayer potential. To see that the lines separating regions of distinct topological invariants indeed correspond to "permanent metal lines" we calculate the band gap using the low energy Hamiltonian. The eigenvalues of this Hamiltonian are

\begin{equation}
E = \pm \sqrt{(\Delta - \abs{k})^2 + \abs{O}^2}
\end{equation}
and so the gap minimum is at \(\abs{k} = \Delta\), and can only vanish on this circle for

\begin{equation}
O = \frac12 (V_\mathrm{AB} e^{\mathrm{i}\phi} - V_\mathrm{BA}e^{-\mathrm{i}\phi}) = 0.
\end{equation}
Rewriting the potentials in polar form $a=0$ implies

\begin{equation}
\abs{V_\mathrm{AB}} e^{\mathrm{i}(\phi + \theta_\mathrm{AB})} = \abs{V_\mathrm{BA}}e^{-\mathrm{i}(\phi - \theta_\mathrm{BA})} 
\label{gapc}
\end{equation}
and so for the band gap to vanish at some momentum angle \(\phi\) a necessary and sufficient condition is thus that two potentials have the same magnitude, \(\abs{V_\mathrm{AB}} = \abs{V_\mathrm{BA}}\).
The two complex numbers on either side of the equality are then identical for \(\phi= (\theta_\mathrm{BA}-\theta_\mathrm{AB})/2\), from which $O=0$ follows.
To determine which stacking vectors this corresponds to, we employ the stacking potentials in the first star approximation which from Eq.~\eqref{interlayer_coupling} are found to be

\begin{align}
V_\mathrm{AB}&=t^{(0)}\left[1+2e^{-\frac{2\pi\mathrm{i}}{a}u_x}\cos\left(\frac{2\pi}{\sqrt{3}a}u_y \right)\right]\label{stackPot2}\\
V_\mathrm{BA}&=t^{(0)}\left[1+2e^{-\frac{2\pi\mathrm{i}}{a}u_x}\cos\left(\frac{2\pi}{\sqrt{3}a}u_y+\frac{2\pi}{3}\right)\right]
\label{stackPot3}
\end{align}
and insertion of these potentials in Eq.~\eqref{gapc} then yields

\begin{align}
u_y&=\frac{\sqrt{3}a}{2} n+\frac{a}{\sqrt{3}}\label{metalic1}\\
u_y &= \pm \sqrt{3} a u_x + \sqrt{3}am + \frac{a}{\sqrt{3}}
\label{metalic23}
\end{align}
with \(m,n \in \mathbb Z\). This corresponds precisely to the three lines on which the winding number changes sign.

\section{Numerical method}

\begin{figure*}
\centering
\includegraphics[width=0.85\textwidth]{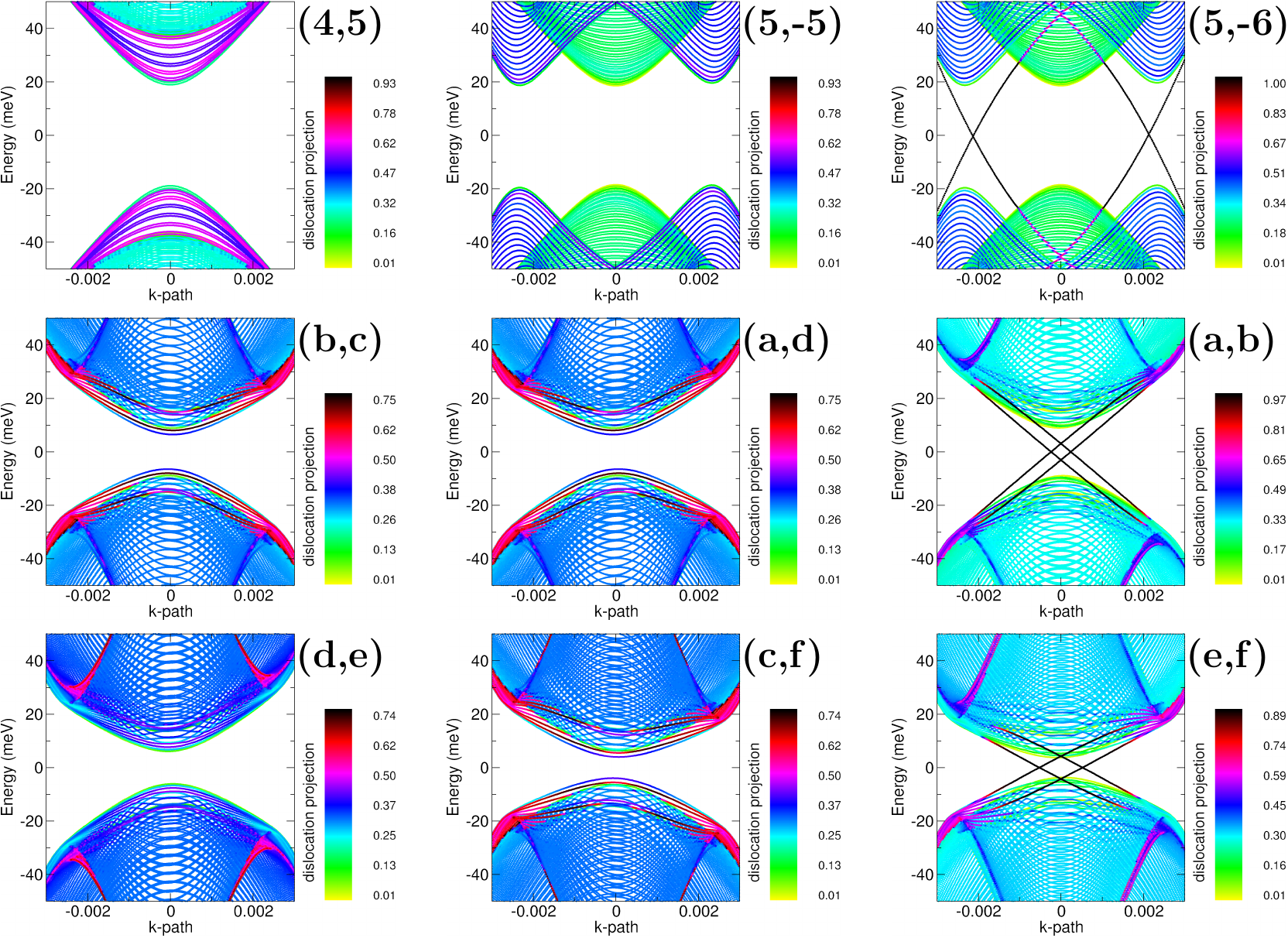}
\caption{Probing the topological phase diagram of bilayer graphene. Band structures for domain walls created between stacking types close to the "metal lines" of the stacking phase diagram separating  regions of different valley Chern numbers. The labeling of each panel corresponds to the stacking vector either side of the domain wall, as indicated in Fig.~\ref{fig1}, with the two stacking regions connected by a continuous change in stacking ("domain wall") given by Eq.~\eqref{tanh} ($w=50a$). The band structures in each row are almost identical, but in the third column gapless states appear. As can be seen from the phase diagram Fig.~\ref{fig1}, in this column the stacking vectors fall either side of the metal line, and hence have different valley Chern numbers, while in the first two columns the stacking vectors fall on the same side of the metal line.}
\label{fig2}
\end{figure*}

In order to probe both the veracity and consequences of the general relation between topological index and stacking order, Eq.~\eqref{C}, we now turn to numerical calculations. In what follows we describe our methodology for both electronic structure simulation and atomic relaxation.

\subsection{Electronic structure calculations}

We model the interlayer displacement field between the regions with stacking vectors $\bu_{1,2}$ by

\begin{equation}
\Delta \bu = \bu_1 + (\bu_2-\bu_1)\tanh(\frac{L(x-x_0)}{w})
\label{tanh}
\end{equation}
that depends on three parameters; the location of the boundary \(x_0\), its width \(w\) and the length of the unit unit cell \(L\). As we employ periodic boundary conditions, we require two domain boundaries which we locate at $x_0 = 1/3$ and $x_0 = 2/3$.
For our tight-binding calculations we employ a \(\pi\)-band only approximation and take the in-plane and interlayer hopping functions to be parameterized by the same Gaussian form

\begin{figure}
\centering
\includegraphics[width=0.50\textwidth]{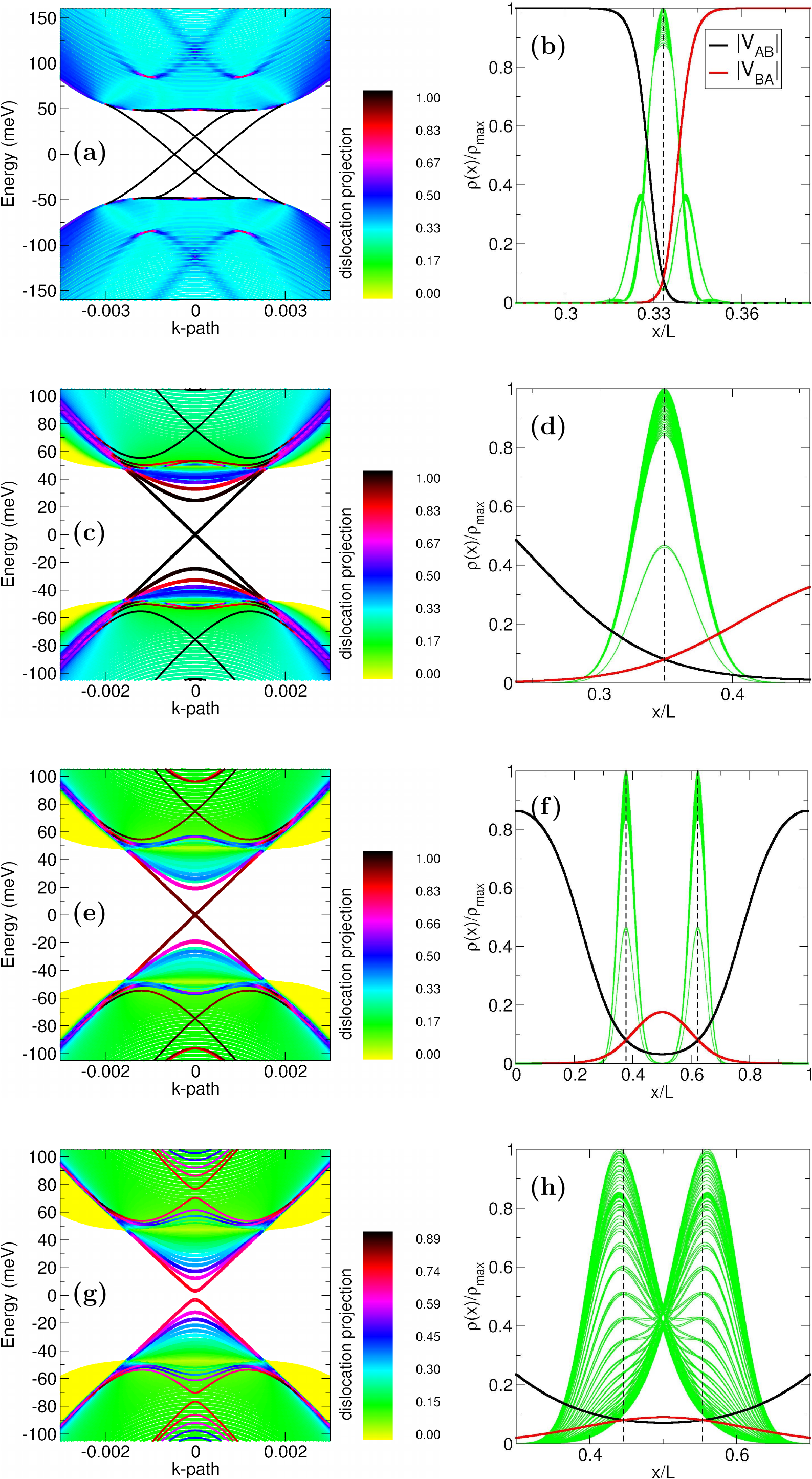}
\caption{Bound states at the crossing of $V_{AB}$ and $V_{BA}$ potentials. {\it Left hand panels}: Band structures for unit cells containing two boundaries (at $x/L = 1/3$, and $2/3$) with increasing dislocation widths of \(50a\), \(1000a\), \(1250a\) and \(1500a\) for panels (a), (c), (e), and (g) respectively. The length of the unit cell is $L = 10000a$. {\it Right hand panels}: The red and black lines indicate the corresponding $V_{AB}$ and $V_{BA}$ potentials. In panel (b) the potentials indicate a sharp domain wall connecting regions of AB and BA stacking, with a smooth stacking modulation as $w$ increases, panel (f). The green lines are the square of the wavefunction for all states in the gap of the pristine bilayer (100~meV), see panel (a). Despite the increasingly smooth modulation, in all cases at the crossing of the $V_{AB}$ and $V_{BA}$ potentials is seen a series of pronounced bound states. (Note in panels (b), (d) and (h) a restricted region of the unit cell is shown for ease of identification of the correspondence of potential crossing with bound state).}
\label{fig3}
\end{figure}

\begin{equation}
t({\v \delta}) = Ae^{-B {\v \delta}^2}
\label{gauss}
\end{equation}
with \(A_\parallel\) and \(B_\parallel\) are chosen to give an in-plane nearest neighbor hopping of 2.8\(\:\)eV and the interlayer \(A_\perp\) and \(B_\perp\) chosen such that the hopping between nearest interlayer neighbours in the AB structure is 0.4\(\:\)eV. The magnitude of $B_\perp$ determines how fast the interlayer interaction decays.

For numerical work we do not need to enforce a restriction to linear momentum (Dirac-Weyl approximation) and instead use a Hamiltonian in which the layer diagonal blocks are the full tight-binding description, with the layer off-diagonal blocks treated through Eq.~\eqref{interlayer_coupling}:

\begin{equation}
H=\begin{pmatrix}
H_\mathrm{TB}^{(1)}&S(x)\\
S^\dagger(x) & H_\mathrm{TB}^{(2)}
\end{pmatrix}
\label{bilHam}
\end{equation}
It is numerically efficient to use a basis of single layer eigenstates, determined from the layer diagonal blocks as\cite{fleischmann_perfect_2020}

\begin{equation}
H_\mathrm{TB}^{(n)}\ket{\Psi_{i\v k}^{(n)}} = \epsilon_{i \v k}^{(n)}\ket{\Psi_{i\v k}^{(n)}}
\end{equation}
We find a basis size of 1600 of the lowest energy states from each layer provides good convergence for the low energy electronic structure of Eq.~\eqref{bilHam}. In this basis the matrix elements of Eq.~\eqref{bilHam} are given by 

\begin{equation}
[H]_{n'i'\v k ' ni\v k}=\delta_{n'i'\v k'ni\v k}\epsilon_{i \v k}^{(n)} + (1-\delta_{nn'})\mel{\Psi_{i'\v k'}^{(n')}}{S(x)}{\Psi_{i\v k}^{(n)}}
\end{equation}

\subsection{Lattice relaxation}

To calculate atomic relaxation we employ the GAFF force field \cite{GAFF} for the C--C interactions within the graphene layers and the registry-dependent interlayer potential of Kolmogorov-Crespi \cite{KC2005} using our own implementation\cite{butz14,fleischmann_perfect_2020}.
For the ideal AB-stacked graphene bilayer this calculational setup results in an equilibrium lattice constant of $a_0=2.441$\,{\AA} and an interlayer distance of $d_{\rm AB}=3.370$\,{\AA}. Shifting the graphene layers to AA stacking increases the layer separation to $d_{\rm AA}=3.597$\,{\AA} ($+0.227\si{\angstrom}$ as compared to AB stacking). The AA-stacked bilayer has a higher energy of 4.4\,meV per atom as compared to AB-stacking, corresponding to a stacking fault energy of $\gamma_{\rm AA} = 54.9$\,mJ/m$^2$. In SP stacking order (see Fig.~2) the equilibrium distance of the graphene layers and the stacking fault energy are $d_{\rm SP}=3.390$\,{\AA} (+0.020\,\AA) and $\gamma_{\rm SP} = 7.1$\,mJ/m$^2$ (0.6\,meV per atom), respectively, in excellent agreement with ACFDT-RPA calculations of Srolovitz \emph{et al.} \cite{sor15}.

\section{Probing the phase diagram}
\label{numbers}

\begin{figure*}
\centering
\includegraphics[width=0.99\textwidth]{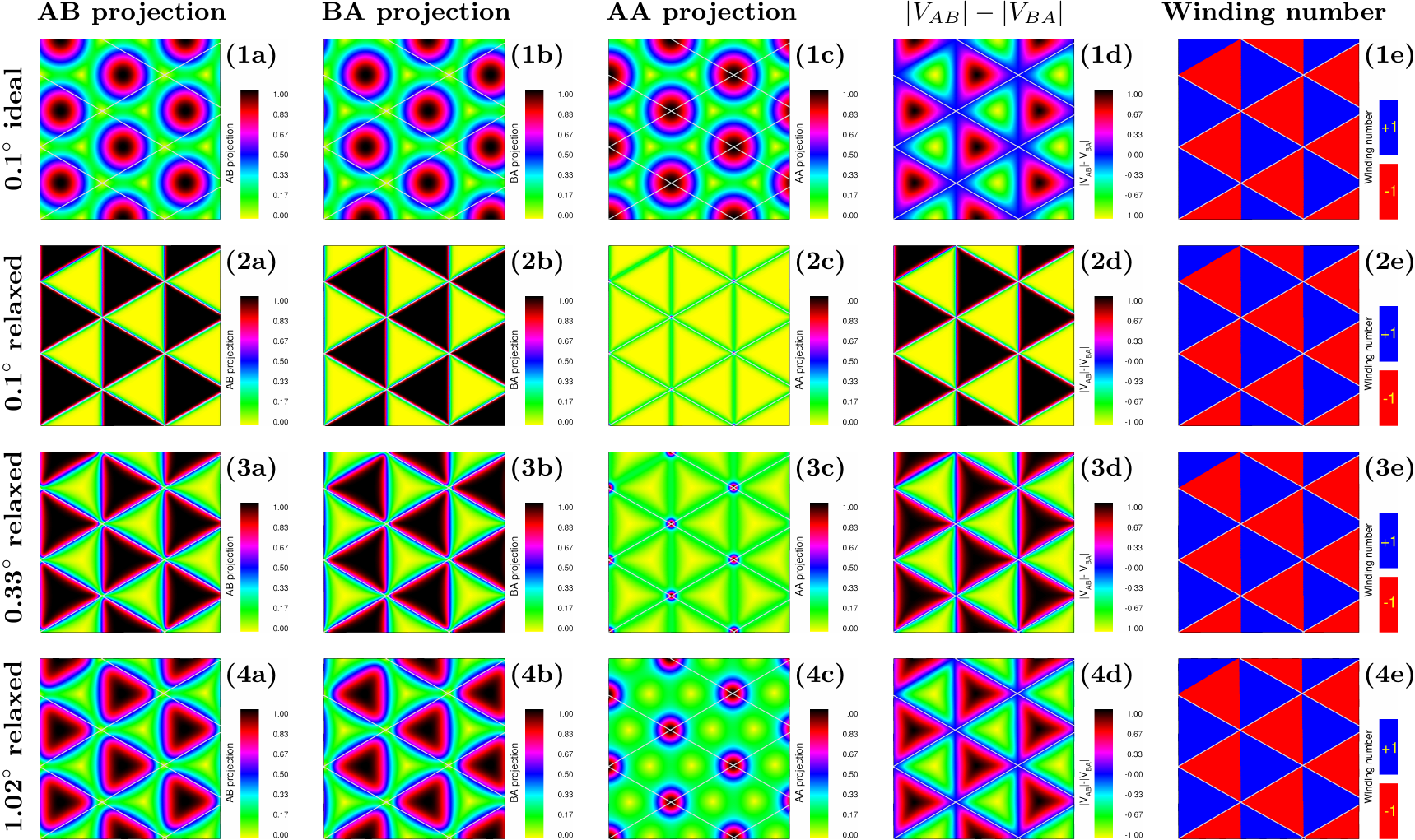}
\caption{Stacking order and Chern index of the twist bilayer. In columns (a-c) are shown the absolute values of the $V_{AB}$, $V_{BA}$, and $V_{AA}$ potentials of a series of twist bilayers. The ideal twist bilayer for $\theta = 0.1^\circ$ ($[p,q]=[1,661]$ in the notation of Ref.~\onlinecite{shall13}) is shown in panels 1a-1c, and a similar picture would be found for any angle. In rows 2-4 the stacking potentials for the relaxed twist bilayer at twist angles of $0.1^\circ$, $0.33^\circ$, ($[p,q]=[1,199]$) and $1.02^\circ$ ($[p,q]=[1,65]$) are displayed. Evidently, at small angles the ideal and relaxed structures appear to be two completely different materials. The difference of $|V_{AB}|-|V_{BA}|$, however, exhibits a closer resemblance between the different systems (panels 1d-4d), and the sign of this difference, which determines the valley Chern number, is essentially identical for all systems (panels 1e-4e).}
\label{fig4}
\end{figure*}

We consider a model system consisting a periodic unit cell with domain walls at $x=1/3$ and $x=2/3$ separating regions of stacking in the sequence $\bu_1 \to \bu_2 \to \bu_1$. By choosing $\bu_{1}$ and $\bu_{2}$ to have either the same or different valley Chern numbers according to the phase diagram of Fig.~\ref{fig1}, a robust test of the relation between valley Chern number and stacking vector can be performed. In Fig.~\ref{fig2} we show band structures that result from choosing $\bu_{1,2}$ in this way. The corresponding stacking vectors for each panel can be read off from the panel label and, as may be observed, in each case where $\bu_{1,2}$ fall either side of a metal line gapless states are found in the spectrum.

As a further numerical test, we probe the occurrence of bound states at nodes of the  function \(\text{sign}\left(\abs{V_\mathrm{AB}}-\abs{V_\mathrm{BA}}\right)\), in systems with smoothly modulated stacking potentials. We consider a unit cell $L=10000a$ in which we have two partial dislocations at $x=1/3$ and $x=2/3$ separating regions of AB and BA stacking; the stacking sequence through the unit cell is thus AB$\to$BA$\to$AB, with the domain walls characterized by the partial Burgers vectors $(0,-\sqrt{2}/3)a$ and $(0,+\sqrt{2}/3)a$. A smooth stacking variation can then be obtained simply by allowing the partial width to become comparable to $L$. In Fig.~\ref{fig3} we see the band structures (left column) and squared wave functions and interlayer potentials (right column) for dislocation widths of \(w=50a\) (a realistic patial dislocation width), and \(1000a\), \(1250a\) and \(1500a\). For the systems shown here we have taken $B_\perp = 4$, a fast decaying potential. This implies that for the misregistry of the layers seen within the core of partial dislocation a weak interlayer interaction, as for hopping vectors much greater in length than the minimal interlayer nearest neighbour separation the hopping matrix element quickly falls to zero. This is the reason for the overall weaker interaction seen at the centre of the cell. While this decay is significantly faster than in bilayer graphene (the tight-binding fitting of Ref.~\onlinecite{lee_zero-line_2016}
 corresponds to $B_\perp = 0.43$) it generates the crossing of AB and BA potentials that we require for a numerical test of Eq.~\eqref{C}. 

As can be seen from Fig.~\ref{fig3}, as the sharp AB$\to$BA$\to$AB transition of the partial dislocation is broadened to a smooth modulation, an increasing number of states appear in the gap, which is almost closed for the $w=1250a$ system. However, for each system there are two crossing points of the stacking potentials $|V_{AB}|$ and $|V_{BA}|$ and at each, as predicted by the change in topological index, bound states are seen in the right hand panel. (Note that to clearly associate the bound state with the crossing of $|V_{AB}|$ and $|V_{BA}|$ in panels (b), (d), and (h), we show a restricted view of a single crossing point.) For the $50a$ domain wall two right moving and two left moving linear gapless states very similar to those reported in the literature\cite{lee_zero-line_2016} can be seen; this is expected from bulk boundary correspondence as the difference in Chern number across the boundary is 2. As the dislocation broadens the gap fills with an increasing number of additional states. In each case, however, exactly at the crossing points of $|V_{AB}|-|V_{BA}|$ bound states are observed, fulfilling the expectation of Eq.~\eqref{C}. Note that all states in the gap of the pristine bilayer (100~meV) are shown in right hand panels, which accounts for the large number of states in each panel.

\section{A Chern index map of the twist bilayer}

\begin{figure*}
\centering
\includegraphics[width=0.69\textwidth]{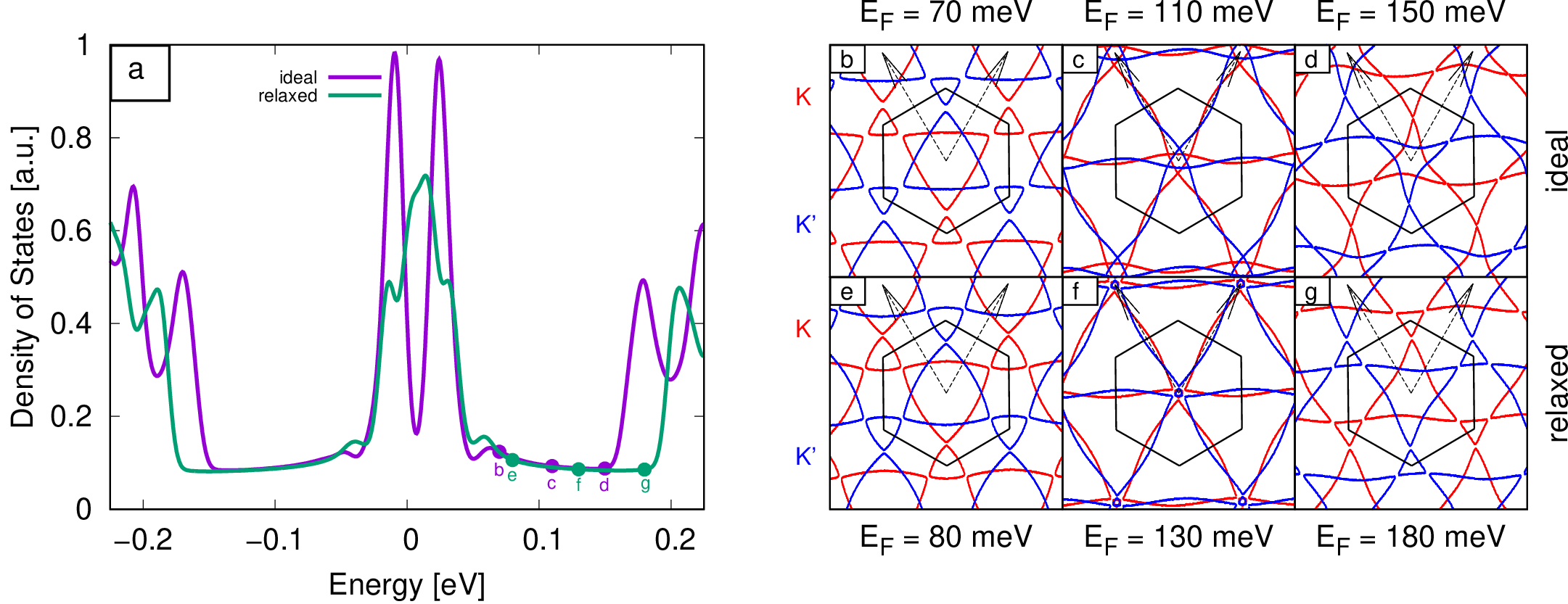}
\caption{Density of states and Fermi surfaces for relaxed and unrelaxed minimally twisted bilayer graphene for a twist angle of $\theta=1.02^\circ$ in an applied bias of $V=200$~meV. {\it Left hand panel}: While the density of states changes significantly close to the Dirac point upon atomic relaxation, the low DOS region between the Dirac point and shoulder peaks remains very similar. {\it Right hand panel}: This robustness to relaxation extends to the Fermi surfaces, which upon relaxation exhibit hybridization at the intersection of the nested Fermi lines and some change in the nesting vector, but remain qualitatively the same in both ideal and relaxed structures. Note that the energies at which the Fermi surfaces are evaluated are scaled so that they correspond to the same relative position between the Dirac and shoulder peaks.}
\label{fig5}
\end{figure*}

Having numerically tested the veracity of the relation between valley Chern number and stacking vector, we now address the question as to how the spatial variation of the valley Chern index is impacted by lattice relaxation in minimally twisted bilayer graphene. For the ideal twist bilayer (row 1, $\theta=0.1^\circ$) the stacking potentials show an equal weight of AB, BA and AA stacking types in the system (as they would for any three stacking projections which would produce a very similar picture but with potential maxima shifted off the high symmetry positions). Upon lattice relaxation this potential landscape dramatically alters: in row 2 we see that the AA potential has all but vanished, remaining only weakly visible at the dislocation core and nodes, with the AB and BA potentials describing a mosaic tiling with $C_3$ symmetry. Increasing the twist angle smooths the edges of this mosaic, and increases strength of the AA potential contribution, see row 3 ($\theta=0.33^\circ$) and row 4 ($\theta=1.02^\circ$). The 4th and 5th columns of this figure display the difference $|V_{AB}|-|V_{BA}|$, and the winding number $\mathrm{sign}(|V_{AB}|-|V_{BA}|)$. Remarkably, we see that the spatial variation of the winding number is identical for all systems. From the results of Secs.~\ref{valleyChernSection} and \ref{numbers}, this indicates that the formation of valley-momentum helical states, which is driven by the changing valley Chern number, will be impacted only in details by lattice relaxation.

To examine this we show in Fig.~\ref{fig5} the density of states and Fermi surfaces for a twist bilayer of $\theta=1.02^\circ$ ($[p,q]=[1,65]$ in the notation of Ref.~\onlinecite{shall13}). While the density of states shows pronounced changes close to the Dirac point, the "valley" between the Dirac point peak and the two shoulder peaks remain largely unchanged. This low, almost constant DOS in the valley region corresponds to the gapless topological states, and as can be seen from the Fermi surfaces, Fig.~\ref{fig5}b, the details of this band structure remain qualitatively the same, with some increased hybridization due to relaxation opening the intersection points of the nested Fermi surface (particularly seen in panel Fig.~\ref{fig5}g).

\section{Discussion}

We have provided a general relation between the topological invariant of bilayer graphene and the stacking vector that describes mutual translation of the layers. We find that the Chern index is given by $C = \nu - \text{sign}(|V_{AB}|-|V_{BA}|)$, with $|V_{AB}|$ and $|V_{BA}|$ the AB and BA components of the interlayer stacking potential. This generalizes the well known result that AB and BA stacked bilayer graphene have valley Chern numbers of 0 and 2 (for the K and K' valley) and 2 and 0 respectively. A consequence of this generalization is that the valley Chern number is now associated with a condition on the interlayer fields rather than the fixed AB and BA structures, and this allows consideration of the occurrence of topologically protected states in regions of smooth stacking variation, such as moir\'es. As a numerical test of this we have performed simulations of artificially broadened domain walls finding bound states at the crossing of the $|V_{AB}|$ and $|V_{BA}|$, as would be expected due to the change in value of $C$ at this point.

With this tool in hand we have examined the valley Chern number for minimally twisted bilayer graphene, finding that the underlying spatial dependence of the valley Chern index is, essentially, independent of atomic relaxation. The topological physics of this material, in particular helical network states, is thus qualitatively similar in the ideal and relaxed twist bilayer. In fact, the ideal twist geometry can be expected to have a much "cleaner" manifestation of the helical network due to the reduced scattering as the interlayer interaction contains only three (first star) momentum boosts, as opposed to the continuum of momentum boosts of the dislocation network.

\subsection*{Acknowledgments}

The work was carried out in the framework of the SFB 953 “Synthetic
Carbon Allotropes” (project number 182849149) of the Deutsche
Forschungsgemeinschaft (DFG). F.W. thanks the Graduate School
GRK 2423 (DFG, project number 377472739) for financial support.
S.S. and S.S. acknowledge the funding of the DFG, reference
SH 498/4-1.


\end{document}